\documentclass[intlimits,twoside,a4paper]{article}

\usepackage{amsmath,amssymb}
\usepackage{graphicx,color}

\usepackage[T2A]{fontenc}
\usepackage[cp1251]{inputenc}

\usepackage[eqsecnum]{cmpj2}

\usepackage{amssymb,amsfonts,amscd}
\usepackage{amsmath,euscript}

\newcommand\blue[1]{\textcolor{blue}{#1}}

\newcommand{\cref}[1]{\blue{\ref{#1}}}

\issue{2015}{18}{3}{33705}
\doinumber{10.5488/CMP.18.33705}

\title[Vibronic interaction in crystals with the Jahn-Teller centers]%
{Vibronic interaction in crystals with the Jahn-Teller centers in the elementary energy bands concept}
\author[D.M. Bercha \textsl{et al.}]
{D.M. Bercha\refaddr{a1}, S.A. Bercha\refaddr{a1}, K.E. Glukhov\refaddr{a1}, M. Sznajder\refaddr{a2}}
\addresses{
\addr{a1} Institute of Physics and Chemistry of Solid State, Uzhgorod National University,
\\ 54 Voloshin St.,  88000 Uzhgorod, Ukraine
\addr{a2}Faculty of Mathematics and Natural Sciences, University of Rzesz\'ow, Pigonia 1, 35-959
Rzesz\'ow, Poland
}

\authorcopyright{D.M. Bercha, S.A. Bercha, K.E. Glukhov, M. Sznajder, 2015}
\date{Received April 6, 2015, in  final form June 3, 2015}

\begin{document}

\maketitle

\begin{abstract}
The order-disorder type phase transition caused by the vibronic interaction (collective Jahn-Teller effect) in a monoclinic CuInP$_2$S$_6$ crystal is analyzed. For this purpose, a trigonal protostructure model of CuInP$_2$S$_6$ is created, through a slight change in the crystal lattice parameters of the CuInP$_2$S$_6$ paraelectric phase. In parallel to the group-theoretical analysis, the DFT-based {\it ab initio} band structure calculations of the CuInP$_2$S$_6$ protostructure, para-, and ferriphases are performed. Using the elementary energy bands concept, a part of the band structure from the vicinity of the forbidden energy gap, which is created by the $d$-electron states of copper, has been related with a certain Wyckoff position where the Jahn-Teller's centers are localized. A construction procedure of the vibronic potential energy matrix is generalized for the case of crystal using the elementary energy bands concept and the group theoretical method of invariants. The procedure is illustrated by the creation of the adiabatic potentials of the $\Gamma_5$--$\Gamma_5$ vibronic coupling for the protostructure and paraphase of the CuInP$_2$S$_6$ crystal. A structure of the obtained adiabatic potentials is analyzed, followed by conclusions on their transformation under a phase transition and the discussion on the possibility for the spontaneous polarization to arise in this crystal.

\pacs 73.21.Cd, 33.20.Wr, 71.70.Ej, 71.15.Mb

\end{abstract}


\section*{Introduction}\label{par_Intro}

The order-disorder type phase transitions occur in a series of compounds, including the CuInP$_2$S$_6$ one \cite{FPCIPS} and, in particular in the case of this crystal, they can be explained  by the realization of the cooperative Jahn-Teller effect (see e.g., references \cite{Maisonneuve_1,JT_CIPS}). The theory of this effect described in literature \cite{Bersuker} is based on the partially model approach. Namely, the effect of vibronic interaction in the Jahn-Teller's centers in a unit cell of a crystal is considered, and further, the interaction between these centers is modeled.
It should be emphasized, that a procedure describing the realization of the collective Jahn-Teller effect is discussed in book \cite{Bersuker}. However, as opposed to the consideration here, it is not founded on the real band structure of crystal. Instead, the collective Jahn-Teller effect is studied in reference \cite{Bersuker} in a model way, i.e., the effect is analyzed initially at the isolated structural unit that coincides with the Jahn-Teller center (unit cell of a crystal or another atoms formation which has a degenerated electronic state and is unstable with respect to Jahn-Teller effect), and next, minimization of the potential energy of interaction between structural units is performed.


As opposed to molecules, where the degenerate or pseudodegenerate local electron states take part in the vibronic process, in crystals the band structure $E(\bf k)$ over the Brillouin zone (BZ) should be taken into account. This circumstance forces us into searching for a new approach to describe the Jahn-Teller effect in crystals which will be the subject of study in this paper. Herein, we present a theory of the Jahn-Teller effect illustrated for the CuInP$_2$S$_6$ crystal. The theory is based on the symmetry of the crystal band structure in the framework of the elementary energy bands (EEBs) concept, introduced in papers by Zak \cite{Zak80, Zak82}. The essence of this concept is that the information about the symmetry and topology of the band structure of a crystal is encoded in the site-symmetry group of a certain Wyckoff position, that was identified later on as the actual Wyckoff position
\cite{YAP,Glukhov07}. The physical meaning of the actual Wyckoff position has been demonstrated in our papers \cite{YAP, Glukhov07}, where it has been shown that the maximum of the spatial valence electron density distribution is focused in this position in the unit cell of a crystal. Moreover, representations of the irreducible band representation describing the symmetry of the EEBs that form the crystal valence band
can be induced only from the irreducible representations (irrep) of the site-symmetry group of the actual Wyckoff position. The most evident example of a relation between the EEB's symmetry, actual Wyckoff position, and the localization of maximum of the valence spatial density distribution in this position are germanium, silicon, A$_3$B$_5$ type crystals, and the superlattices based on them \cite{Glukhov07}.

It should be expected that the band structure of crystals with the Jahn-Teller centers will be composed of the EEBs reflecting the local symmetry of certain Wyckoff positions which, in turn, coincide with certain Jahn-Teller centers. Since the EEBs concept allows us to present the `spatial issue' (i.e., the energy spectrum of crystal) as the issue concerning a point symmetry, it is possible to utilize in our approach the theory of the Jahn-Teller effect, elaborated for molecules \cite{Jahn-Teller}.

The structure of this paper is as follows. In section~\ref{par_Struct}, the information on the  crystalline structure of CuInP$_2$S$_6$ and its phase transition related to the Jahn-Teller effect is presented. A low symmetry of the CuInP$_2$S$_6$ paraphase permits one to consider only the cooperative Jahn-Teller effect \cite{Bersuker}. In order to demonstrate the existence of nearby-in-energy local electronic levels which are necessary to discuss this effect, a modelling of the high-symmetry CuInP$_2$S$_6$ `protostructure' is performed in section~\ref{par_PraPhase}, together with the comparative group-theoretical analysis of energy states of all CuInP$_2$S$_6$ phases. Section~\ref{par_AbInitio} presents the results of the DFT-based {\it ab initio} band structure calculations of all phases of the CuInP$_2$S$_6$ crystal. Attention is paid to the presence of the EEB in the band structure of paraphase, which is related to the Wyckoff position $d\left(\frac{2}{3},\frac{1}{3},\frac{1}{4}\right)$ where Cu atom is located. In section~\ref{par_JT}, the theory of Jahn-Teller effect is formulated, together with its generalization for the case of a crystal. A special role of the actual Wyckoff position as the information carrier about the electronic component of the vibronic instability is explained. Additionally, the symmetry of normal vibrations which are active in the Jahn-Teller effect is discussed for the CuInP$_2$S$_6$ protostructure, based upon the irreducible representations of the site-symmetry group of the actual Wyckoff position. In the vibronic instability analysis, the normal vibrations which are associated with the degenerate states near the energy gap have been checked with respect to their activities in the above mentioned group-theory sense, for several high-symmetry points in the BZ. This has been done to confirm the validity of using the site-symmetry group of the actual Wyckoff position, whose symmetry encodes the symmetry of the EEB associated with $d$-states of copper. Section~\ref{par_AdiabPot} is devoted to the construction of the vibronic potential energy in a matrix form, as well as of the adiabatic potentials for protostructure and paraphase. The Pikus' method of invariants \cite{Bir74}  is used there for the first time to solve such kind of a problem. The final section~\ref{par_Concl} presents the analysis on the completeness of our approach in the description of the Jahn-Teller effect, as a mechanism of the transition from the para- to ferriphase in the CuInP$_2$S$_6$ crystal.

\section{Structure and symmetry of the CuInP$_2$S$_6$ crystal}\label{par_Struct}

\begin{figure}[!t]
\centering
\includegraphics[width=0.95\textwidth]{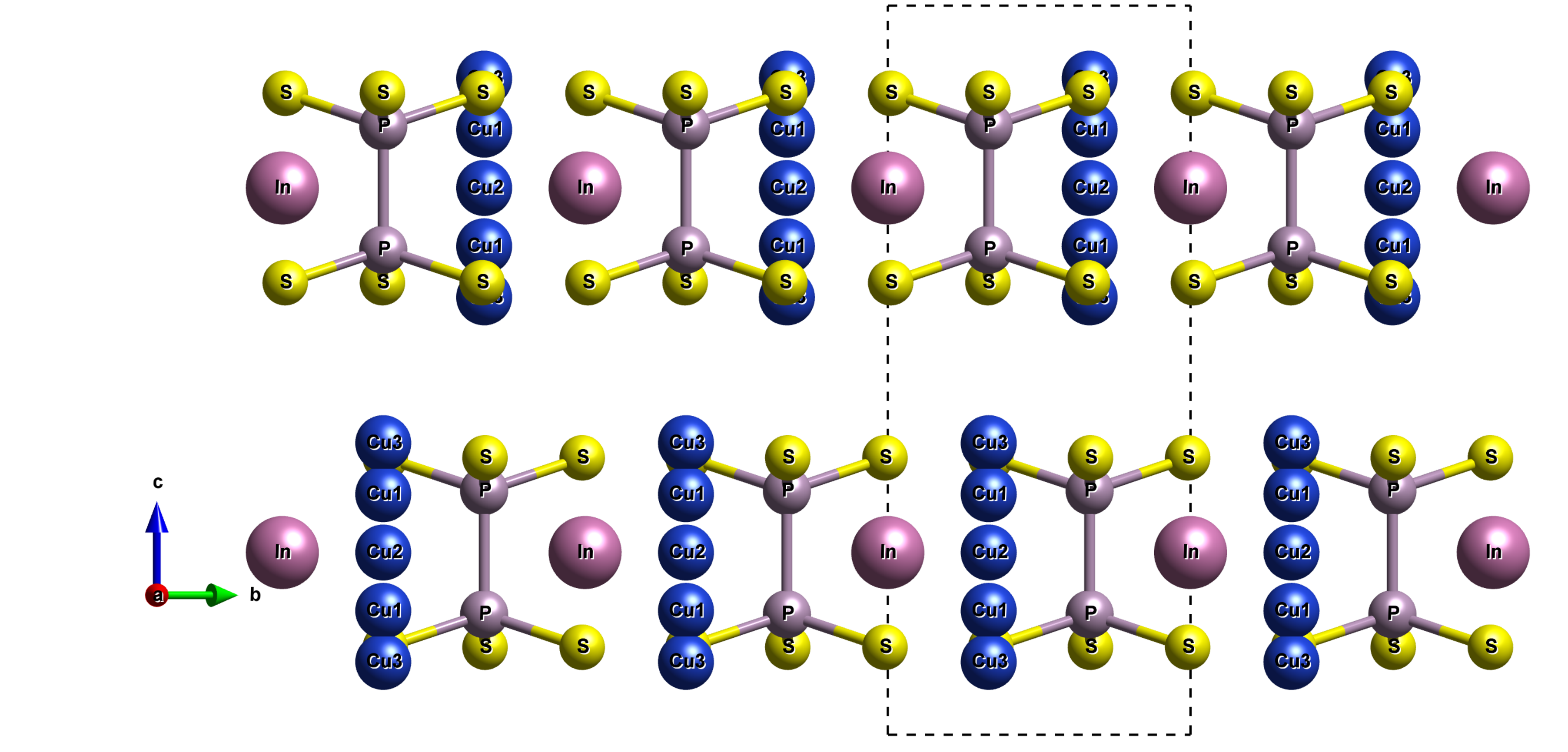}
 \caption{(Color online) Projection of the CuInP$_2$S$_6$ crystal structure. Dashed line encompasses the primitive unit cell of the protocrystal. }\label{fig1}
\end{figure}

The CuInP$_2$S$_6$ crystal possesses a layered structure (see figure~\ref{fig1}) with the atomic layers  separated by the van der Waals gaps \cite{CIPS}. A single atomic layer is composed of the octahedral sulfur framework in which Cu, In, and P--P atom pairs fill the octahedral voids. A peculiarity of the CuInP$_2$S$_6$ crystal structure is the presence of three types of copper atoms sites which are partially occupied, i.e., (i)  quasitrigonal Cu$1$, shifted from the centers of the octahedra, (ii) octahedral Cu$2$, located in the centers of the octahedra, and (iii) nearly tetrahedral Cu$3$, which penetrates into the interlayer space. The occupancy of these positions varies significantly with temperature \cite{Maisonneuve_1}. Furthermore, there are two types of Cu$1$ positions: Cu$1^\textrm{u}$ which is shifted upward from the middle of the layer, and Cu$1^\textrm{d}$, shifted downwards from it. At low temperatures  ($T<153$~K), copper atoms fully occupy positions Cu$1^\textrm{u}$ \cite{FPCIPS}. Upon heating, the occupancy of Cu$1^\textrm{u}$ position decreases while the occupancy of Cu$1^\textrm{d}$ begins to increase. A hopping motion between  Cu$1^\textrm{u}$ and Cu$1^\textrm{d}$ positions leads to an increase in the atomic layer thickness, i.e., to an increase in the volume of the elementary unit cell without change of the number of structural units \cite{FPCIPS}.

In the region of the phase transition from the ferrielectric to paraelectric phase ($T_\textrm{c}=315$~K), the positions Cu$1^\textrm{u}$ and Cu$1^\textrm{d}$ are filled with equal probability and the polarity of both copper sublattices disappears. This phase transition is accompanied by the space symmetry group change, within the monoclinic system, from Cc (C$_\textrm{s}^4$)  (ferriphase) to C2/c (C$_\textrm{2h}^6$) (paraphase). Above $T_\textrm{c}=315$~K, the Cu$1^\textrm{u}$ and Cu$1^\textrm{d}$ sites become equivalent, while at higher temperatures the Cu$1$ occupancy decreases, and the Cu$2$ and Cu$3$ sites start to fill up (see figure~\ref{fig1}). In our study below, we shall focus on the premises of the order-disorder type phase transition that occurs at $T_\textrm{c}=315$~K.

\begin{figure}[!b]
  \centering
\includegraphics[width=0.35\textwidth]{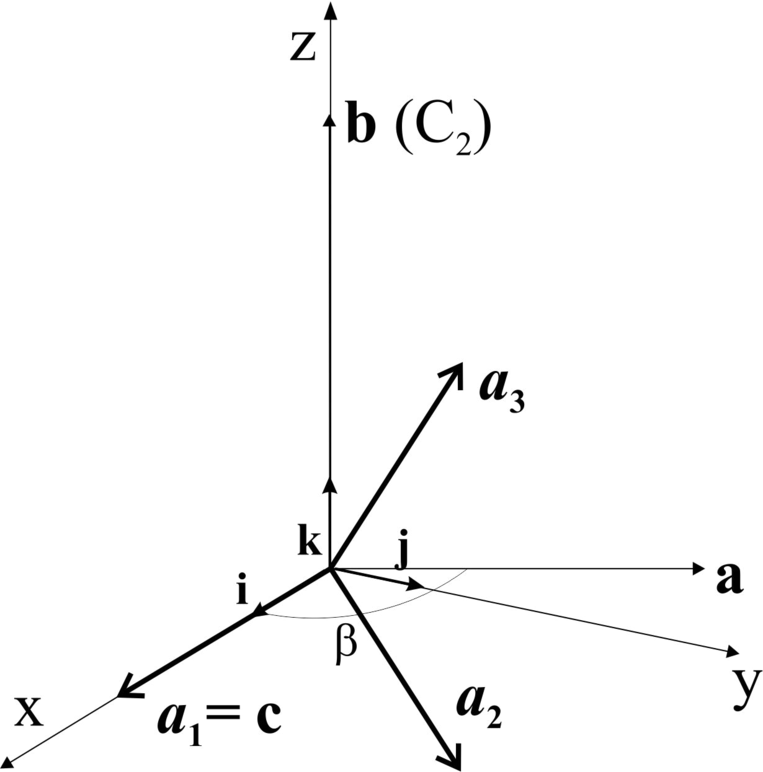}
  \caption{Basis vectors of the base-centered monoclinic primitive lattice of the CuInP$_2$S$_6$ paraphase in the non-orthogonal $x$, $y$, $z$ coordinate system.}\label{fig2}
\end{figure}

The lattice parameters of both ferrielectric and paraelectric phases of CuInP$_2$S$_6$ are as follows \cite{Maisonneuve_1}: $a=6.09559$~{\AA}, $b=10.56450$~{\AA}, $c=13.6230$~{\AA}, $\beta=107.1011^\circ$, while the primitive cell parameters are $a_1=13.6230$~{\AA}, $a_2=a_3=6.096846$~{\AA}, $\alpha=120.0311^\circ$, $\beta=\gamma=98.4508^\circ$. Crystals of both phases belong to the monoclinic base-centered lattice. The basis vectors of the primitive lattice can be expressed by the lattice parameters in the following way ${\boldsymbol a}_1={\bf c}$, ${\boldsymbol a}_2=({\bf a-b})/2$, ${\boldsymbol a}_3=({\bf a+b})/2$, and we associate with them a
non-orthogonal $x, y, z$ coordinate system shown in figure~\ref{fig2}.

The symmetry group C$_\textrm{2h}^6$ (C12/c1) of the CuInP$_2$S$_6$ paraphase is chosen in such a way that the two-fold leading axis coincides with vector {\bf b}. Moreover, the values of its lattice parameters allow us to construct a hexagonal cell of the CuInP$_2$S$_6$ protostructure with {\bf c} axis directed along the basis vector {\bf a} of the monoclinic lattice, using only a small displacements of atoms. To accomplish this task we have first calculated the atomic positions of the CuInP$_2$S$_6$ paraphase by symmetrizing those of the ferriphase, and preserving the same lattice
parameters. The obtained atomic coordinates are presented in table~\ref{tab_struct}. The model of the CuInP$_2$S$_6$ protostructure will be discussed in the next section.

\begin{table}[!t]
  \centering
  \caption{Atomic coordinates of the CuInP$_2$S$_6$ para- and ferriphases.}\label{tab_struct}
 \vspace{2ex}
  \begin{tabular}{|c|c|c|c|}
    \hline
Structure & Coordinates & Site & Site-symmetry group \\ \hline\hline
paraphase
&   Cu $(0.5000,0.8355,0.2500)$ & $4e$ & $2$ \\
&   In $(0.5000,0.5019,0.2500)$ & $4e$ & $2$ \\
&    P $(0.5591,0.1682,0.1193)$ &  $8f$ & $1$\\
&S$_1$ $(0.7296,0.1620,0.1193)$ & $8f$ & $1$\\
&S$_2$ $(0.7612,0.8304,0.1237)$ & $8f$ & $1$ \\
&S$_3$ $(0.2439,0.0117,0.1215)$ & $8f$ & $1$\\ 
ferriphase~\cite{CIPS}
& Cu$^\textrm{u}$ $(0.5957,0.8355,0.3869)$ & $4a$ & $1$ \\
& Cu$^\textrm{d}$ $(0.4310,0.8355,0.1490)$ & $4a$ & $1$ \\
& In $(0.5000,0.5019,0.2500)$ & $4a$ & $1$ \\
& P$1$ $(0.5686,0.1690,0.3491)$ &  $4a$ & $1$\\
& P$2$ $(0.4505,0.1674,0.1788)$ &  $4a$ & $1$\\
& S$1$ $(0.2808,0.1512,0.3950)$ & $4a$ & $1$\\
& S$2$ $(0.2332,0.1645,0.8930)$ & $4a$ & $1$ \\
& S$3$ $(0.7845,0.0177,0.3950)$ & $4a$ & $1$\\
& S$4$ $(0.7400,0.1727,0.1336)$ & $4a$ & $1$\\
& S$5$ $(0.7555,0.1747,0.6404)$ & $4a$ & $1$\\
& S$6$ $(0.2722,0.9943,0.6379)$ & $4a$ & $1$\\ \hline
  \end{tabular}
\end{table}
\noindent

\section{Model of CuInP$_2$S$_6$ protostructure and comparative group-theoretical analysis}\label{par_PraPhase}

A paraphase of the CuInP$_2$Se$_6$ crystal which is relative to CuInP$_2$S$_6$ belongs to the trigonal system with hexagonal Bravais lattice and is described by the D$_\textrm{3d}^2$ space symmetry group \cite{CIPSe}. Both this fact and the appropriate values of the CuInP$_2$S$_6$ primitive lattice parameters ($\boldsymbol a_1, \boldsymbol a_2, \boldsymbol a_3, \alpha, \beta, \gamma$) urge us to create a model of the CuInP$_2$S$_6$ protostructure with the trigonal symmetry. This can be done in the following way. A three-fold leading axis can be directed along the primitive vector $\boldsymbol a_1$, which will become the basis vector $\bf c$ in the hexagonal unit cell, ${\boldsymbol a_1}=\bf c$. A slight deformation applied to the angles $\beta$ and $\gamma$ can transform them into the right angles. As a result, a hexagonal lattice is obtained with parameters $c=13.623$~{\AA} and the angle $\gamma= 120^\circ$, which roughly coincides with the angle $\alpha= 120.0311^\circ$ of the monoclinic unit cell. Hence, the monoclinic $\boldsymbol a_1$ axis becomes the $\bf c$ axis of the hexagonal lattice, and two monoclinic basis vectors with length $a_2 = a_3$ spanning the angle $120^\circ$ lie in the plane perpendicular to $\bf c$ axis. Next, the atomic coordinates of the CuInP$_2$S$_6$ paraphase are slightly changed from their original sites in such a way that the obtained trigonal CuInP$_2$S$_6$ protostructure is described by the D$_\textrm{3d}^2$ (P$\bar{3}$12/c) space group of the relative CuInP$_2$Se$_6$ crystal. The atomic coordinates of the CuInP$_2$S$_6$ protostructure are contained in table~\ref{tab_protostruct}.
\begin{table}[!h]
  \centering
  \caption{The atomic coordinates of the trigonal model of CuInP$_2$S$_6$ protostructure.}\label{tab_protostruct}
 \vspace{2ex}
    \begin{tabular}{|c|c|c|c|}
      \hline
Atom & Coordinates & Site \;\; & Site-symmetry group \\ \hline\hline
Cu &   $(\frac23,\frac13,\frac14)$ & $2d$ & $3\;.\;2$ \\
In &   $(0,0,\frac14)$ & $2a$ & $3\;.\;2$ \\
P  &   $(\frac13,\frac23,0.1655)$ &  $4f$ & $3\;.\;.$\\
S  &   $(0.3306,0.3401,0.1201)$ & $12i$ & $1$\\ \hline
  \end{tabular}
\end{table}

\begin{figure}[!t]
\centering
\includegraphics[width=0.65\textwidth]{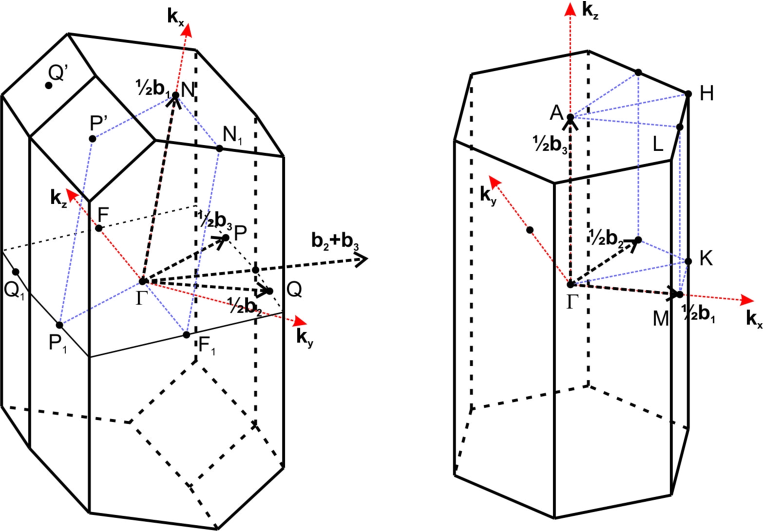}
  \caption{(Color online) Left: Brillouin zone for the para- and ferrielectric phases of the monoclinic CuInP$_2$S$_6$ crystal and its high symmetry points, $\Gamma$: 0, $N$: $\frac12{\bf b}_1$, $Q$: $\frac12{\bf b}_2$, $P$: $\frac12{\bf b}_3$, $F$: $\frac12({\bf b}_3-{\bf b}_2)$,
$N_1$: $\frac12[{\bf b}_1 + ({\bf b}_2-{\bf b}_3)]$, $Q'$: $\frac12({\bf b}_1 - {\bf b}_2)$, $P'$: $\frac12({\bf b}_1 - {\bf b}_3)$. Right: hexagonal BZ for the model protostructure. High-symmetry points: $\Gamma$: 0, $M$: $\frac12{\bf b}_1$, $A$: $\frac12{\bf b}_3$, $L$: $\frac12({\bf b}_1+{\bf b}_3)$, $K$: $\frac13({\bf b}_1+{\bf b}_2)$,
$H$: $\frac13({\bf b}_1+{\bf b}_2)+\frac12{\bf b}_3$, $P$: $\frac13({\bf b}_1+{\bf b}_2)+\mu{\bf b}_3$.}\label{fig3}
\end{figure}

As it is known, there exist 6 different kinds of the Brillouin zone for a monoclinic lattice \cite{Bir74}, depending on the relation between the basis vectors length of the direct and the reciprocal spaces, respectively. In our case, the BZ is presented in figure~\ref{fig3}, together with the description of the high-symmetry points, expressed by the combination of ${\bf b}_1$, ${\bf b}_2$, and ${\bf b}_3$ reciprocal lattice vectors \cite{Bir74}. Passing from the monoclinic to the hexagonal model cell. these three vectors are transformed in the following way, ${\bf b}_1\rightarrow {\bf b}_3^\textrm{hex}$, ${\bf b}_2\rightarrow {\bf b}_2^\textrm{hex}$, ${\bf b}_3\rightarrow {\bf b}_1^\textrm{hex}$, the angle $\angle ({\bf b}_1^\textrm{hex}, {\bf b}_3^\textrm{hex})= \angle ({\bf b}_2^\textrm{hex}, {\bf b}_3^\textrm{hex})=90^\circ$, while $\angle ({\bf b}_1^\textrm{hex}, {\bf b}_2^\textrm{hex})=60^\circ$. The resulting hexagonal BZ with high-symmetry points of the CuInP$_2$S$_6$ protostructure is presented in the right-hand part of figure~\ref{fig3}. A correspondence between the high-symmetry points, their irreducible representations of the wave vector groups for the trigonal CuInP$_2$S$_6$ protostructure and the CuInP$_2$S$_6$ para- and ferriphases is displayed in table~\ref{tabl2}.

The investigation on the additional degeneracy of energy states due to the time reversal symmetry for some high-symmetry points of the BZ, on the presence of extrema in $E(\bf k)$ dependencies in these points, as well as on the dispersion laws near these points have been performed for the symmetry groups C$_\textrm{2h}^6$ (paraphase) and C$_\textrm{s}^4$ (ferriphase) in paper \cite{Bercha78}. In order to determine the changes in the energy spectrum parameters when passing from the protostructure to the CuInP$_2$S$_6$ paraphase, we have performed a group-theoretical analysis of the $E(\bf k)$ dependencies in some high-symmetry points of the hexagonal BZ (D$_\textrm{3d}^2$ space symmetry group). This comparative group-theoretical analysis will serve as an additional verification of the validity to simulate the phase transition by using the CuInP$_2$S$_6$ protostructure model. In our studies the Herring's criterion \cite{Herring}, Rashba formula \cite{Rashba}, as well as the Pikus' method of invariants \cite{Bir74} have been applied. Table~\ref{tab4} presents the investigation results concerning the presence of extrema in the $E(\bf k)$ dependencies in certain directions of the hexagonal and monoclinic BZs, as well as of the additional degeneracy of energy states due to the time reversal symmetry, which is classified according to the Herring's criterion. Comparing the results of paper \cite{Bercha78} and Table IV it can be stated that when passing from the protostructure to the CuInP$_2$S$_6$ paraphase, neither changes in the presence of additional degeneracy of states at most high-symmetry points, nor in the presence of the $E(\bf k)$ extrema in particular directions of the reciprocal space can be observed.

In order to solve the issue concerning the relationship between the energy spectrum and the instability of the system caused by the vibronic interaction, attention should be paid to the following high-symmetry points of the hexagonal BZ $\Gamma(0,0,0)$, $A(0,0,\frac12)$, $M(\frac12,0,0)$, $L(\frac12,0,\frac12)$, $H(\frac13,\frac13,\frac12)$, $K(\frac13,\frac13,0)$, as well as to their counterparts from the BZ of a monoclinic crystal, as will be shown below. From table~\ref{tab4} it follows that the extrema in the $E(\bf k)$ dependencies can be observed in all three main directions of the BZ in the vicinity of certain high-symmetry points of hexagonal BZ and their counterparts, i.e.,
$ \Gamma \,(\Gamma)$, $A \,(N)$, $L \,(P',Q')$, $M \,(P,Q)$. The discussed D$_\textrm{3d}^2$ space symmetry group exhibits 6 irreducible representations in the BZ center \cite{Kovalev}, two of which ($\Gamma_5\, (E_g)$ and $\Gamma_6\,(E_u)$ in notation by Kovalev \cite{Kovalev}) are two-dimensional. Therefore, they are of particular interest for our study. These representations are reduced to two one-dimensional representations, describing split energy states, at the transition from C$_\textrm{2h}^6$ (paraphase) to C$_\textrm{s}^4$ (ferriphase). This is an important remark that validates the necessity to model the trigonal protostructure of the CuInP$_2$S$_6$ crystal. A near-in-energy distance between the split energy states, described by the one-dimensional irreducible representations, can indicate slight changes in structural parameters when passing form the real paraphase to the trigonal CuInP$_2$S$_6$ protostructure. As will be shown below, the {\it ab inito} band structure calculation results confirm both the presence of a small splitting of the $\Gamma_5$ and $\Gamma_6$ degenerate states when passing form the trigonal to monoclinic CuInP$_2$S$_6$ crystal, and the occurrence of such splitting at other BZ points of the monoclinic lattice.

\begin{table}[!t]
  \centering
  \caption{The correspondence between the high-symmetry points, as well as between the irreducible representations for the  hexagonal BZ of protostructure and the monoclinic para- and ferriphases of CuInP$_2$S$_6$.}\label{tabl2}
  \vspace{2ex}
  \begin{tabular}{|c|c|c|c|}
\hline
Hexagonal BZ\;\; & Irrep \;\; & Monoclinic BZ & Irrep
  \\ \hline\hline
  $\Gamma(0,0,0)$ &
  \begin{tabular}{c}
  $\Gamma_1$ \\
  $\Gamma_2$ \\
  $\Gamma_3$ \\
  $\Gamma_4$ \\
  $\Gamma_5$ \\
  $\Gamma_6$ \\
  \end{tabular}
  & $\Gamma(0,0,0)$ &
  \begin{tabular}{c}
  $\Gamma_{1}$ \\
  $\Gamma_{2}$ \\
  $\Gamma_{3}$ \\
  $\Gamma_{4}$ \\
  $\Gamma_{1}+\Gamma_{3}$ \\
  $\Gamma_{2}+\Gamma_{4}$ \\
  \end{tabular}
  \\  
~ & ~ & ~ & ~ \\
  $A(0,0,\frac12)$ &
  \begin{tabular}{c}
  $A_1$ \\
  $A_2$ \\
  $A_3$ \\
  \end{tabular}
  & $N(\frac12,0,0)$ &
  \begin{tabular}{c}
  $N_1$ \\
  $N_2$ \\
  $N_3$ \\
  \end{tabular}
  \\  
~ & ~ & ~ & ~ \\
  $M(\frac12,0,0)$ &
  \begin{tabular}{c}
  $M_1$ \\
  $M_2$ \\
  $M_3$ \\
  $M_4$ \\
  \end{tabular}
  & $P(0,0,\pm\frac12)$ &
  \begin{tabular}{c}
  $P_1$ \\
  $P_2$ \\
  \end{tabular}
  \\  
~ & ~ & ~ & ~ \\
  $L(\frac12,0,\frac12)$ & $L_1$
  & $P'(\frac12,0,\pm\frac12)$ & $P'_1$\\  
~ & ~ & ~ & ~ \\
  $K(\frac13,\frac13,0)$ &
  \begin{tabular}{c}
  $K_1$ \\
  $K_2$ \\
  $K_3$ \\
  \end{tabular}
  & $V(\mu_1,\mu_2,\mu_2)$ &
  \begin{tabular}{c}
  $V_1$ \\
  $V_2$ \\
  $V_1+V_2$ \\
  \end{tabular}
  \\  
~ & ~ & ~ & ~ \\
  $H(\frac13,\frac13,\frac12)$ &
    \begin{tabular}{c}
  $H_1$ \\
  $H_2$ \\
  $H_3$ \\
  \end{tabular}
  & $V(\mu_1,\mu_2,\mu_2)$ &
  \begin{tabular}{c}
  $V_1$ \\
  $V_2$ \\
  $V_1+V_2$ \\
  \end{tabular}
  \\  
~ & ~ & ~ & ~ \\
  $P(\frac13,\frac13,\mu)$ &
    \begin{tabular}{c}
  $P_1$ \\
  $P_2$ \\
  $P_3$ \\
  \end{tabular}
  & $V(\mu_1,\mu_2,\mu_2)$ &
  \begin{tabular}{c}
  $V_1$ \\
  $V_2$ \\
  $V_1+V_2$ \\
  \end{tabular}
  \\  \hline
\end{tabular}
\end{table}

\begin{table}[!t]
  \centering
  \caption{The effect of the time-reversal symmetry on the presence of band extrema in the main directions of the hexagonal and monoclinic BZs of CuInP$_2$S$_6$, as well as on the additional degeneracy of representations expressed by the Herring's criterion (cases $a$ or $b$).  The local coordinate systems have the same orientation of axes for all high symmetry points in both BZs.}\label{tab4}
  \vspace{2ex}
  \begin{tabular}{|c|c|} \hline 
  Hexagonal & Monoclinic \\ 
  \begin{tabular}{cccc} 
  Point  &  Irrep  &  Case  &  $\frac{\partial E_n}{\partial k_i}=0$ \\ \hline
  $\Gamma$ &
  \begin{tabular}{c}
  $\Gamma_1$ \\
  $\Gamma_2$ \\
  $\Gamma_3$ \\
  $\Gamma_4$ \\
  $\Gamma_5$ \\
  $\Gamma_6$ \\
  \end{tabular}
  & $a_1$ & $i=x,y,z$ \\ 
~& ~& ~& ~ \\
  $A$ &
  \begin{tabular}{c}
  $A_3$ \\
  $\{A_1+A_2\}$ \\
  ~\\
  \end{tabular}
  &
  \begin{tabular}{c}
  $a_1$ \\
  $b_1$ \\
    ~\\
  \end{tabular} &
    \begin{tabular}{c}
  $i=x,y$ \\
  $-$ \\
    ~ \\
  \end{tabular} \\ 
~& ~& ~& ~ \\
    $M$ &
  \begin{tabular}{c}
  $M_1$ \\
  $M_2$ \\
  $M_3$ \\
  $M_4$ \\
  \end{tabular}
  & $a_1$ & $i=x,y,z$ \\ 
~& ~& ~& ~ \\
  $L$ &
  \begin{tabular}{c}
  $L_1$ \\
  \end{tabular}
  & $a_1$ & $i=x,y,z$ \\ 
~& ~& ~& ~ \\
  $K$ &
  \begin{tabular}{c}
  $K_1$ \\
  $K_2$ \\
  $K_3$ \\
  \end{tabular}
  & $a_2$ & $i=x,y,z$ \\ 
~& ~& ~& ~ \\
  $H$ &
  \begin{tabular}{c}
  $H_1$ \\
  $H_2$ \\
  $H_3$ \\
  \end{tabular}
  & $a_2$ & $i=x,y,z$ \\ 
~& ~& ~& ~ \\
  $P$ &
  \begin{tabular}{c}
  $P_1$ \\
  $P_2$ \\
  $P_3$ \\
  \end{tabular}
  & $a_2$ & $i=x,y,z$ \\
  \end{tabular} &
  \begin{tabular}{cccc} 
  Point  &  Irrep   &  Case  &  $\frac{\partial E_n}{\partial k_i}=0$  \\ \hline
  $\Gamma$ &
  \begin{tabular}{c}
  $\Gamma_1$ \\
  $\Gamma_2$ \\
  $\Gamma_3$ \\
  $\Gamma_4$ \\
  $\Gamma_1+\Gamma_3$ \\
  $\Gamma_2+\Gamma_4$ \\
  \end{tabular}
  & $a_1$ & $i=x,y,z$ \\ 
~& ~& ~& ~ \\
  $N$ &
  \begin{tabular}{c}
  $N_1$ \\
  $N_2$ \\
  $N_3$ \\
  \end{tabular}
  &
  \begin{tabular}{c}
  $a_1$ \\
  \end{tabular} &
    \begin{tabular}{c}
  $i=x,y$ \\
  \end{tabular} \\ 
~& ~& ~& ~ \\
  $P$ &
  \begin{tabular}{c}
  ~  \\
  $P_1$ \\
  $P_2$ \\
  ~\\
  \end{tabular}
  & $a_1$ & $i=x,y,z$ \\ 
~& ~& ~& ~ \\
  $P'$ &
  \begin{tabular}{c}
  $P'_1$ \\
  \end{tabular}
  & $a_1$ & $i=x,y,z$ \\ 
~& ~& ~& ~ \\
   $V$ &
  \begin{tabular}{c}
  $V_1$ \\
  $V_2$ \\
  $V_1+V_2$ \\
  \end{tabular}
  & \begin{tabular}{c}
   $a_2$ \\
  \end{tabular} &
    \begin{tabular}{c}
   $i=x,y$ \\
  \end{tabular} \\ 
~& ~& ~& ~ \\
  $V$ &
  \begin{tabular}{c}
  $V_1$ \\
  $V_2$ \\
  $V_1+V_2$ \\
  \end{tabular}
  &\begin{tabular}{c}
  $a_2$ \\
  \end{tabular} &
    \begin{tabular}{c}
  $i=x,y$ \\
  \end{tabular} \\ 
~& ~& ~& ~ \\
  $V$ &
  \begin{tabular}{c}
  $V_1$ \\
  $V_2$ \\
  $V_1+V_2$ \\
  \end{tabular}
  & \begin{tabular}{c}
  $a_2$ \\
  \end{tabular} &
    \begin{tabular}{c}
   $i=x,y$  \\
  \end{tabular} \\
  \end{tabular} \\ \hline
  \end{tabular}
\end{table}

The important information confirming the correct modelling of protostructure at which the parameters of the energy spectrum are changed slightly can be obtained by the analysis of the dispersion laws for charge carriers in the vicinity of certain high-symmetry points from the corresponding Brillouin zones. Below we present the investigation on the dispersion law near the point $\Gamma$ for the doubly-degenerate $\Gamma_5$ state of the CuInP$_2$S$_6$ protostructure.
For this purpose, the Pikus' method of invariants \cite{Bir74} is used. In this method, the secular matrix $D(\bf k)$, which allows one to obtain the $E(\bf k)$ dependence, is presented as a sum of invariants. The invariants are products of the $A_{is}$ basis matrices and the $f(\bf k)$ functions depending on the wave vector components, whose symmetry is defined by means of the symmetric and antisymmetric decomposition of square character of the irreducible representation $\Gamma_5$. The decomposition is defined as~\cite{Bir74}
\begin{equation}
\label{eq1}
n_s^+=\frac{1}{2n}\sum_{g\in G}{\chi^s(g)\left\{[\chi(g)]^2+\chi\left(g^2\right)\right\}}
\end{equation}
and
\begin{equation}
\label{eq2}
n_s^-=\frac{1}{2n}\sum_{g\in G}{\chi^s(g)\left\{[\chi(g)]^2-\chi\left(g^2\right)\right\}},
\end{equation}
where $n_s^+$ and $n_s^-$ denote numbers of  irreducible representations (irreps) $\tau^s$ in the considered symmetric ($n_s^+$) and antisymmetric ($n_s^-$) squares of irreps, $n$ is a number of elements in the space group, the summation runs over elements of the wave vector group, and $\chi^s(g)$ is a character of irreducible representations of the wave vector group in the center of the BZ. From equation (\ref{eq1}) it follows that the even functions $f_+(\bf k)$ are transformed according to the representations $\tau_s=\Gamma_1$, $\Gamma_5$, while from equation (\ref{eq2}), that the uneven functions $f_-(\bf k)$ are transformed according to $\tau_s=\Gamma_3$. The $A_{is}$ basis matrices which enter the sum of invariants can be obtained using a transformation rule of a matrix of the given order, under the effect of symmetry operations. In particular, for $\Gamma_1$, the basis matrix will be identical to the second order unity matrix $\sigma_1$, and for the representation $\Gamma_5$, the basis matrices will be the $\sigma_x$ and $\sigma_z$ Pauli matrices. In order to determine the basis functions $f_+=k_x^2+k_y^2$ (or $f_+=k_z^2$) for the representation $\Gamma_1$,  and $f_+=k_x k_z$ ($k_y k_z$) for $\Gamma_5$, the projection operator method has been used \cite{Bir74}. It should be noted that the obtained $D(\bf k)$ matrix does not contain any components, which are transformed according to the irreducible representation $\Gamma_3$. This is a consequence of parity of the corresponding basis function, whose application result does not coincide with that of an inversion element, for this representation. Our analysis allows us to present the secular matrix $D({\bf k}) $ in the following form:
\begin{equation}\label{eq3}
D\left( {\bf k} \right)=\left( \begin{matrix}
   a\left( k_{x}^{2}+k_{y}^{2} \right)+bk_{z}^{2}+c{{k}_{x}}{{k}_{z}} & c{{k}_{y}}{{k}_{z}}  \\
   c{{k}_{y}}{{k}_{z}} & a\left( k_{x}^{2}+k_{y}^{2} \right)+bk_{z}^{2}-c{{k}_{x}}{{k}_{z}}  \\
\end{matrix} \right).
\end{equation}

By solving the resulting secular equation, the following dispersion law for charge carriers of the CuInP$_2$S$_6$ protostructure is obtained for the electron state described by the irreducible representation $\Gamma_5$,
\begin{equation}\label{eq4}
E\left( {\bf k} \right)=a\left( k_{x}^{2}+k_{y}^{2} \right)+bk_{z}^{2}\pm \sqrt{{{c}^{2}}k_{z}^{2}\left( k_{x}^{2}+k_{y}^{2} \right)}\,.
\end{equation}

Now, it can be checked how the obtained dispersion law is transformed when passing to the CuInP$_2$S$_6$ paraphase. By comparing the characters of irreducible representations of the C$_\textrm{2h}^6$ symmetry group of the monoclinic crystal and the characters of $\Gamma_5$ of D$_\textrm{3d}^2$ (see tables~\ref{tabA1} and \ref{tabA2} in appendix), we conclude that the $\Gamma_5$ representation is reduced to two representations: $\Gamma_1$ and $\Gamma_3$ of the monoclinic crystal.
For small changes in the energy spectrum parameters which are caused by the protostructure model used instead of the CuInP$_2$S$_6$ paraphase, the splitting between the energy states described by the irreducible representations $\Gamma_1$ and $\Gamma_3$ of the monoclinic crystal will be small. Hence, it can be concluded that these states interact with each other. For the joint representations which describe the split interacting states, the $D(\bf k)$ matrix, being the basis for the $E(\bf k)$ dependence, is two-dimensional \cite{Bir74}. The diagonal terms of this matrix are transformed according to the representation $\tau_s = |\Gamma_1|^2 + |\Gamma_3|^ 2 = 2 \Gamma_1$, while the off-diagonal terms, according to $\tau_s = \Gamma_1 \times \Gamma_3 + \Gamma_3 \times \Gamma_1 = 2 \Gamma_3$. As a result, the $D(\vec{k}) $ matrix takes the form:
\begin{equation}\label{eq5}
D\left( {\bf k} \right)=\left( \begin{matrix}
   {{a}_{1}}k_{x}^{2}+{{b}_{1}}k_{y}^{2}+{{c}_{1}}k_{z}^{2}+\frac{\Delta }{2} & \alpha {{k}_{x}}{{k}_{z}}+f{{k}_{y}}{{k}_{z}}  \\
   \alpha {{k}_{x}}{{k}_{z}}+f{{k}_{y}}{{k}_{z}} & {{a}_{2}}k_{x}^{2}+{{b}_{2}}k_{y}^{2}+{{c}_{2}}k_{z}^{2}-\frac{\Delta }{2}  \\
\end{matrix} \right),
\end{equation}
where $\Delta$ denotes the energetic distance between two interacting states. A comparison of equations (\ref{eq3}) and (\ref{eq5}) shows that both matrices are composed of functions of the same wave vector components.
From the analysis presented above it follows that the comparison of the band structures of the paraphase and of the CuInP$_2$S$_6$ protostructure model is essential, in particular, in the $\Gamma$ point. This issue will be discussed in the next section.

\section{\emph{Ab initio} band structure calculations of CuInP$_2$S$_6$ protostructure, \\ para-   and ferriphases}\label{par_AbInitio}

Electronic energy structure of the CuInP$_2$S$_6$ protostructure, para-, and ferriphases has been calculated within the framework of the density functional theory \cite{Kohn64, Kohn65} in the local approximation (LDA) \cite{Ceperley80,Perdew81}, by means of the software packages ABINIT and SIESTA \cite{Gonze02,Soler02}. In our calculations, plane waves and linear combination of atomic orbitals have been used correspondingly as a basis set for ABINIT and SIESTA programs. A periodic crystal structure has been taken into account through the boundary conditions at the boundaries of the unit cell. \emph{Ab initio} norm conserving pseudopotentials \cite{Bachelet82,Hartwigsen98}, for the following electron configurations of atoms have been utilized in calculations, Cu: [Ar]~$3d^{10}4s^1$, In: [Kr]~$5s^{2}5p^1$, P: [Ne]~$3s^{2}3p^3$, and S: [Ne]~$3s^{2}3p^4$.

The cutoff energy  $E_\textrm{cut} = 20$~Ry of plane waves for the self-consistent calculation has been chosen to obtain a convergence in the total energy of the cell not worse than 0.001~Ry/atom. Such basis set consists of about 6000 plane waves. The total and partial electronic densities of states have been determined by a modified method of tetrahedra \cite{Bloechl94}, for which the energy spectrum and wave functions are calculated on the 80 points $k$-mesh. Integration over the irreducible part of the Brillouin zone has been performed using the special $k$-points method \cite{Chadi73,Monkhorst76}. Finally, the optimization of the structural parameters has been done for all phases CuInP$_2$S$_6$. A contribution of particular atomic orbitals in the creation of
various valence band ranges has been analyzed by means of the partial density of states function.

\begin{figure}[!t]
  \centering
\includegraphics[width=0.7\textwidth]{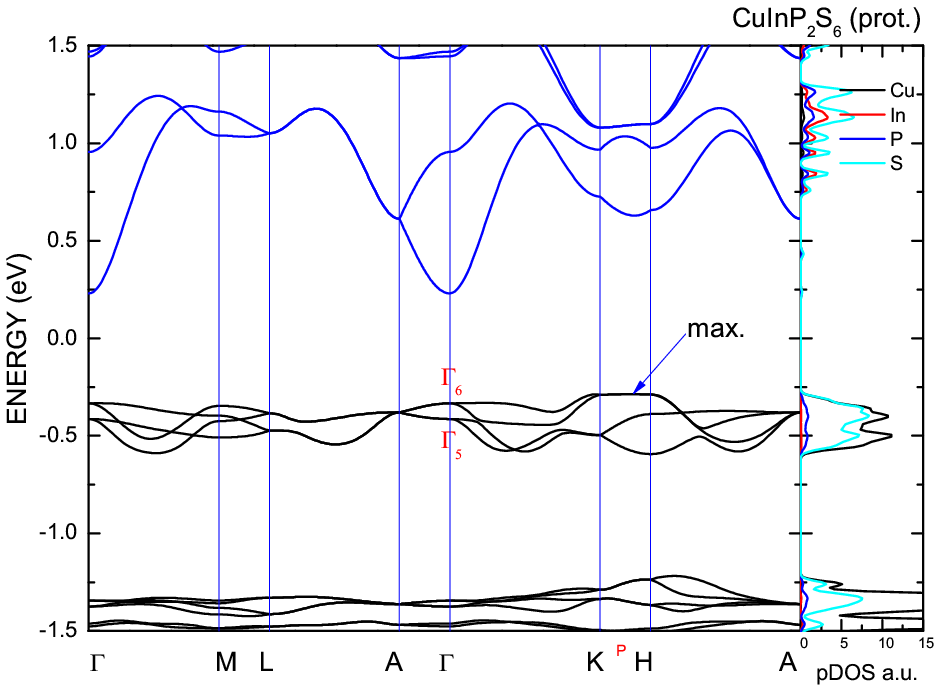}
  \caption{(Color online) \emph{Ab initio} band structure and density of states of the CuInP$_2$S$_6$ trigonal protostructure. The calculated valence band maximum is in $K$--$H$ (or $P$) direction. Black line in the right-hand panel indicates the contribution of Cu atoms in the total DOS.}\label{fig4}
\end{figure}

\begin{figure}[!b]
  \centering
\includegraphics[width=0.7\textwidth]{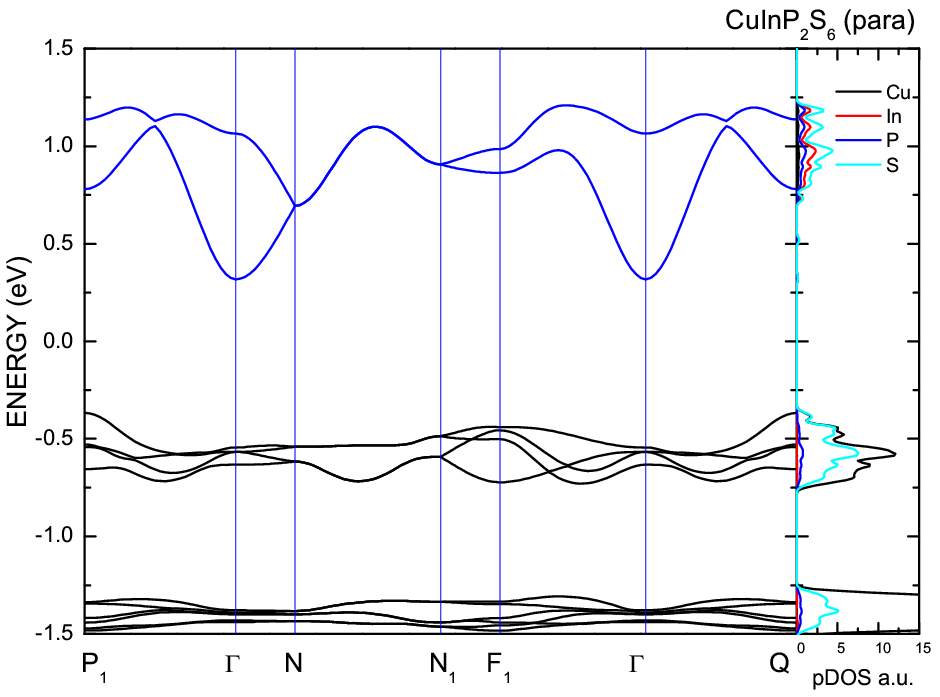}
  \caption{(Color online) \emph{Ab initio} band structure and density of states of the CuInP$_2$S$_6$ paraelectric phase. }\label{fig5}
\end{figure}

Figures~\ref{fig4} and \ref{fig5} show parts of the band spectrum of the protostructure and paraphase of CuInP$_2$S$_6$ crystal, respectively, together with the partial density of electronic states functions. As can be seen, the obtained valence band edges throughout the Brillouin zones are weakly dispersive for both phases of CuInP$_2$S$_6$. Both phases exhibit indirect bandgaps. In particular, the valence band top of the protostructure is located in the $H$--$K$ direction, while the conduction band minimum is present in the $\Gamma$ point.

\begin{figure}[!h]
  \centering
\includegraphics[width=0.7\textwidth]{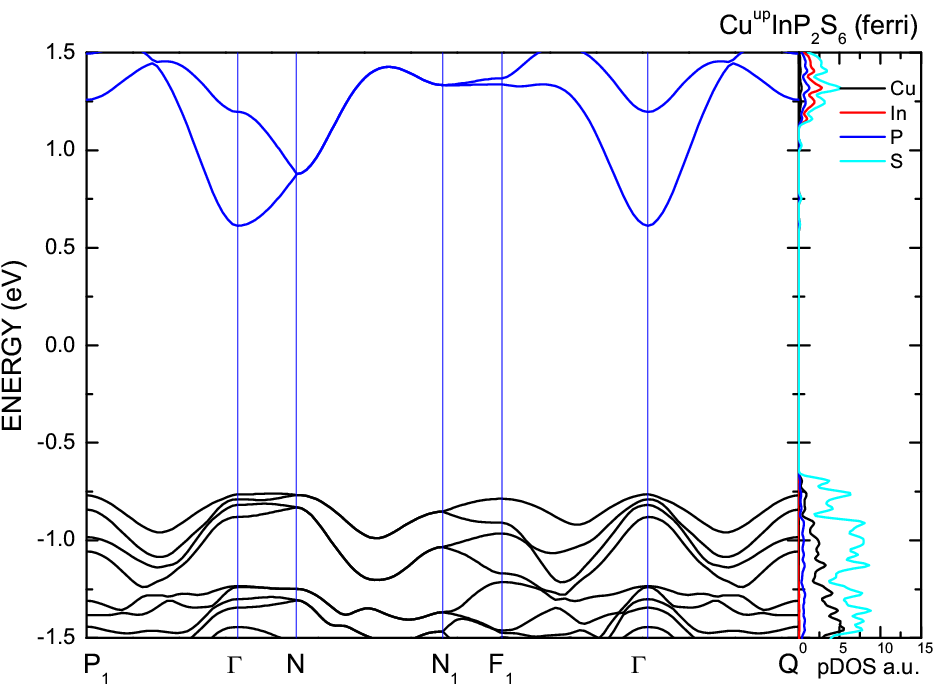}
  \caption{(Color online) \emph{Ab initio} band structure and density of states of the CuInP$_2$S$_6$ ferrielectric phase. Configuration of copper site: Cu$^\textrm{u}$.}\label{fig6}
\end{figure}

Moreover, the valence band top of the CuInP$_2$S$_6$ paraphase (figure~\ref{fig5}) is composed of the elementary energy band which consists of four weakly split branches. Since the valence band top of the CuInP$_2$S$_6$ protostructure exhibits an identical topology (i.e., 4-branch EEB), it can be stated that this elementary energy band is indeed suitable for analysis of protostructure~--- paraphase transformation of the CuInP$_2$S$_6$. In addition to the band structure calculations, we have performed the symmetry description of the obtained energy states near the energy gap in the $\Gamma$ point of both phases. As a result, two highest energy states of the CuInP$_2$S$_6$ protostructure are described by two-dimensional irreducible representations $\Gamma_5$ and $\Gamma_6$ in the $\Gamma$ point, in the direction of increasing energies. By comparing figures~\ref{fig4} and \ref{fig5} it can be seen that these states undergo splitting when passing from the trigonal protostructure to the monoclinic paraphase. As a result, four nearby-in energy states described by one-dimensional irreducible representations from the monoclinic C$_\textrm{2h}^6$ space group are obtained. Furthermore, the representations are reduced in the following way, $\Gamma_5 \rightarrow \Gamma_2 +\Gamma_4$, $\Gamma_6 \rightarrow \Gamma_1 +\Gamma_3$. The obtained representations exhibit different parity, i.e., $\Gamma_2$ and $\Gamma_4$ are uneven representations, while $\Gamma_1$ and $\Gamma_3$ are even ones.
When the symmetry of the system is further lowered, i.e., the transition from paraelectric to ferrielectric phase occurs, the discussed four states become more distant in energy and the corresponding four-branch subband becomes more spread which indicates that the connection between the respective branches is weaker (see figure~\ref{fig6}).

\section{Vibronic interaction as a mechanism of the configuration change of molecule and of the order-disorder phase transition in crystals with degenerate electron states}\label{par_JT}

As it is known, some symmetric molecules can undergo the Jahn-Teller effect \cite{Jahn-Teller}, which consists in lowering the symmetry of molecule due to the electron-vibronic interaction. As a result of this interaction, a degenerate electronic term is split and a rearrangement of the vibronic spectrum occurs. For the molecule configuration to be stable, the molecule energy, which is a function of the distance between cores, should exhibit a minimum at the given configuration of cores. It means that the expression that describes the energy change due to a small shift of core positions cannot contain terms linear with respect to the normal coordinates. Neglecting the adiabatic approximation, the expression that describes the potential energy of an electron subsystem can contain terms dependent on the electron subsystem coordinates. This can lead to the instability of the molecule, change in its configuration, as well as to the degeneracy removal of the electronic state, which takes part in this vibronic interaction.

A part of the Hamiltonian of electron subsystem related to the deviation from the adiabatic approximation can be written with accuracy to quadratic terms of the normal coordinates, which are treated as parameters of the potential energy, in the following way:
\begin{equation} \label{eq6}
\hat{H}_1=\sum\limits_{\alpha i}{V_{\alpha i}(r)}{Q_{\alpha i}}+\sum\limits_{\alpha \beta ik}{W_{\alpha \beta ik}(r)Q_{\alpha i}Q_{\beta k}}+\dots\,,
\end{equation}
where $r$ denotes a set of coordinates of the electron subsystem. Expression (\ref{eq6}) is at the same time the perturbation term of the Hamiltonian $H_0$ that describes the electron subsystem of the symmetric configuration of molecules. The first-order perturbation correction linear with respect to normal coordinates, which describes the electronic energy of a molecule, can be written, within the adiabatic approximation by means of the matrix element
\begin{equation}
V_{\rho \sigma }=\sum\limits_{\alpha i}{Q_{\alpha i}}\int{\Psi _{\rho }V_{\alpha i}(r)\Psi_{\sigma}\rd q,}
\end{equation}
where $\Psi_{\rho}$ and $\Psi_{\sigma}$ denote wave functions of the electron subsystem describing a degenerate electron term.

The Hamiltonian $\hat H_1$ is invariant with respect to the transformations of the symmetry group, of considered system. Since it contains a linear term of decomposition in series with respect to $Q_{\alpha i}$, the coefficients $V_{\rho \sigma}(r)$ and $W_{\alpha \beta ik}(r)$, dependent on the coordinates of electron subsystem, are transformed under the action of symmetry elements in the same way as the normal coordinates $Q_{\alpha i}$, or their products. The subscript $\alpha$ denotes the number of irreducible representation, while $i$ is the number of the basis wave function of this representation.

In analogy to the way how the dispersion law $E(\bf k)$ for charge carriers has been found, in our approach a secular equation is used, where the Hamiltonian is presented in a matrix form with the elements $\langle\Psi_{\rho}^* | \hat{H}_1 | \Psi_{\sigma} \rangle $. This Hamiltonian is called the vibronic potential energy operator \cite{Bersuker}. It should be noted that those vibrations which are related to the linear terms of the vibronic interaction operator $\hat H_1$ with respect to the normal coordinates are called active vibrations in the Jahn-Teller effect. By solving the secular equation, the so-called adiabatic potential is obtained, which can be used to predict the presence of some stable or unstable configurations of a molecule.

Recently, a construction procedure of the vibronic potential energy and of the adiabatic potential for a high-symmetry molecule by means of the Pikus' method of invariants has been reported \cite{Sofia13}. It has been demonstrated that the representations $\tau_s$ according to which the functions dependent on the normal coordinate components are transformed, together with the vibrations active in the Jahn-Teller effect, as well as the matrices $A_{is}$ contained in invariants \cite{Bir74}, can be obtained only based upon a decomposition of the symmetric square of character of the irreducible representation, which describes a degenerate electron state.

While investigating the Jahn-Teller effect in a crystal, a problem arises how to transform the criteria obtained for the degenerate electron states to the corresponding energy bands $E(\bf k)$ in the BZ. It is obvious that one should concentrate first on the energy states which form a vicinity of the forbidden energy band gap. In this case, a possibility occurs for exchanging energy between electrons from a degenerate state and the respective phonons, if this process is allowed by the selection rules. It has been demonstrated in the previous section that copper $d$-electron states create a connected so-called elementary energy band throughout the BZ at the valence band top of CuInP$_2$S$_6$. Therefore, as opposed to molecules, one should consider this EEB in the CuInP$_2$S$_6$ crystal, instead of a degenerate $d$-electron level. As it is mentioned in introduction, the symmetry of the EEB is described by the so-called irreducible band representation \cite{Zak80,Zak82}. Since neither periodicity of the CuInP$_2$S$_6$ crystal lattice nor the number of atoms in a unit cell is changed in its phase transition, it should be expected that a phonon with the symmetry described by the representation of the wave vector group in ${\bf q}=0$ will be active in the Jahn-Teller effect. Note that in order to identify a normal mode that is active in the vibronic interaction, we utilize in our study a fact that the EEB of the  CuInP$_2$S$_6$ valence band top originates from the $d$-electron states of copper atoms. These atoms, in turn, occupy the Wyckoff positon $d(\frac13, \frac23, \frac14)$ in the protostructure of CuInP$_2$S$_6$ crystal. Representations of the corresponding irreducible band representation which describes the symmetry of the EEB can be induced from the irreducible representations of the site-symmetry group of one of the Wyckoff position $d$  multiplicities. The coordinates of this position $d(\frac13, \frac23, \frac14)$ coincide, in turn, with the localization of the Jahn-Teller center in the CuInP$_2$S$_6$ crystal. Hence, the actual Wyckoff position $d$ can be regarded as a distinct center, where the information on the electronic band structure of the CuInP$_2$S$_6$ crystal is encoded. Therefore, in order to find the symmetry of a normal vibrational mode that is active in the Jahn-Teller effect, it is enough to consider a decomposition of the symmetric square of characters of the representation induced from the irreducible representations of the site-symmetry group of the actual Wyckoff position $d$. This site-symmetry group coincides with the factor group of space-symmetry group of CuInP$_2$S$_6$ at the point ${\bf k}=0$.

Hence, it can be postulated that the above group-theoretical procedure to find a vibrational mode that is active in the Jahn-Teller effect coincides with the procedure elaborated for molecules. In order to confirm this statement, a direct calculation concerning the Jahn-Teller's criterium can be performed, for all high-symmetry points of the BZ, describing the states from the vicinity of the forbidden energy gap of the discussed crystal. Generally, in the case of centrosymmetrical crystals, the wave vector groups can contain an inversion element $I$. If it is the case, then the action of this element on a wave vector is as follows, $I {\bf k}_0 = -{\bf k}_0 \doteq {\bf k}_0 + {\bf b}$, otherwise $I {\bf k}_0 = -{\bf k}_0 \neq {\bf k}_0 + {\bf b}$, where $\bf b$ is a reciprocal lattice vector. However, the construction of the symmetric square of representation characters differs in both cases. In the first case, the symmetric square of representation characters of the ${\bf k}_0$ wave vector group can be decomposed into irreducible representations of the ${\bf k}_0$ wave vector group by means of the formula  \cite{Rashba}
\begin{equation} \label{eq8}
{{n}_{s}}=\frac{1}{2n}\sum\limits_{g\in {{G}_{{{{\bf k}}_{0}}}}}{\left\{ {{\left[ {{\chi }_{{{{\bf k}}_{0}}}}\left( g \right) \right]}^{2}}+\chi_{{{{\bf k}}_{0}}} \left( {{g}^{2}} \right) \right\} \, {{\chi }_{s}}\left( g \right)},
\end{equation}
where $n_s$ is the number of irreps $\tau^s$ in the considered direct product, $n$ denotes the number of elements in the space group,  and $\chi_s$ is the character of irreducible representations of the wave vector group in the center of the BZ. When the wave vector group does not contain the inversion element, the decomposition is as follows \cite{Rashba},
\begin{equation}\label{eq9}
{{n}_{s}}=\frac{1}{2n}\sum\limits_{g\in {{G}_{{{{\bf k}}_{0}}}}}{\left\{ {{\chi }^{s}}\left( g \right) \, {{\chi }_{{{{\bf k}}_{0}}}}\left( g \right) \, {{\chi }_{{{{\bf k}}_{0}}}}\left( {{R}^{-1}}gR \right)+{{\chi }^{s}}\left( Rg \right){{\chi }_{{{{\bf k}}_{0}}}}\left[ {{(Rg)}^{2}} \right] \right\}},
\end{equation}
where $R$ denotes an element that transforms the ${\bf k}_0$ into $-{\bf k}_0$ (inversion in our case, since it is present in the space group of a centrosymmetrical crystal).

As we have demonstrated in section~\ref{par_AbInitio}, the valence band top of the CuInP$_2$S$_6$ crystal protostructure is composed of the EEB that is created by $d$-electron states of copper. Meanwhile, its symmetry is described by the following irreducible band representation,
\begin{equation}\label{eq10}
\Gamma_5+\Gamma_6 - A_1+A_2 +A_3- M_1+M_2+M_3+M_4 - 2L_1 - K_1+K_2+K_3 - H_1+H_2+H_3\,,
\end{equation}
that is related to the actual Wyckoff position $d(\frac13, \frac23, \frac14)$ in which a copper atom is located. Note that the absolute maximum of the valence band of the CuInP$_2$S$_6$ protostructure is observed in the  $K\frac{~P~}{~}H$ direction. Hence, the analysis of energy states that are situated in these high-symmetry points of the BZ becomes crucial. The wave vector groups at points $K(\frac13, \frac13, 0)$ and $H(\frac13, \frac13, \frac12)$ (see figure~\ref{fig3}, right) do not contain the inversion element. Hence, the decomposition of their symmetric square of representation characters can be described by equation (\ref{eq9}). However, from the viewpoint of the Jahn-Teller effect realization, only two-dimensional representations of the wave-vector groups which are contained in the irreducible band representation (\ref{eq10}) are interesting for study. The wave vector groups of other points in the BZ ($\Gamma(0,0,0)$, $A(0,0,\frac12)$, $M(\frac12,0,0)$, $L(\frac12,0,\frac12)$) contain an inversion element. Hence, the decomposition of their symmetric square of characters of two-dimensional representations (in particular $\Gamma_5$ and $\Gamma_6$) is given by equation (\ref{eq8}). Consequently, representations for the $\Gamma$ point which can be found from equation (\ref{eq8}) become a background to construct the adiabatic potential for the collective Jahn-Teller effect.

\section{Vibronic potential energy and adiabatic potential of CuInP$_2$S$_6$ protostructure and paraphase}\label{par_AdiabPot}

We shall find in the beginning of this section the normal vibrational modes of the CuInP$_2$S$_6$ crystal protostructure which take part in the vibronic instability. By analyzing the transformation of the atomic positions under the action of symmetry elements of the structure, the following expression that describes the character of the mechanical representation is obtained
\begin{equation}\label{eq11}
\chi_M=4\Gamma_1+ 6\Gamma_3+ 10\Gamma_5\,.
\end{equation}
In order to construct a matrix of the vibronic potential energy that is connected with the doubly-dege\-nerate electron states of the EEB (\ref{eq10}), one should establish first the symmetry of the normal vibration (phonon) that is active in the Jahn-Teller effect. Correspondingly, by applying decomposition (\ref{eq8}) to the representations $\Gamma_5$, $\Gamma_6$, and equation (\ref{eq9}) to $K_3$, $H_3$ from the EEB (\ref{eq10}), we obtain the expected result: $n_s \neq 0$, only for $\tau_s = \Gamma_1$, and $\tau_s=\Gamma_5$. It means that these representations can describe the vibrations active in the Jahn-Teller effect. The same result can be found by means of the decomposition of the symmetric square of characters of the irreducible representations $\Gamma_5$ and $\Gamma_6$, from the extended site-symmetry group of the actual Wyckoff position $d$ comprising its two muliplicities $d_1(\frac13, \frac23, \frac14)$ and $d_2(\frac23, \frac13, \frac14)$. In these positions, there are the Jahn-Teller's centers located, i.e., copper atoms. Hence, the normal vibrations that are active in the Jahn-Teller effect exhibit the symmetry described by the irreducible representations $\Gamma_1$, and $\Gamma_5$ in the center of the BZ. However, the vibration $\Gamma_1$ should be excluded from the considerations, since it does not lead to a change in configurations of atoms in a unit cell. Therefore, the matrix of the vibronic interaction potential energy is built based on the normal coordinates $Q_1$ and $Q_2$, being the functions which are transformed according to the representation $\Gamma_5$ of the space group D$_\textrm{3d}^2$. The construction procedure of this matrix is analogous to the way how the dispersion law $E(\bf k)$ was obtained in section~\ref{par_PraPhase}. In order to create the matrix of the vibronic interaction potential energy, it is necessary to construct some invariants in the form of matrices being the products of $Q_1$ and $Q_2$ functions, and their combinations. Theses matrices and functions are transformed according to the representation
$\tau_s=\Gamma_5$. Similarly to the $D(\bf k)$ matrix, the above matrices should be chosen in the form of Pauli matrices, as well as a second-order unity matrix. It is obvious that the normal coordinates  $Q_1$ and $Q_2$ are also transformed according to the representation $\Gamma_5$. Using the projection operator technique and the matrix of representation $\Gamma_5$ written in a real basis (see table~\ref{tabA3} in appendix) one obtains that the functions $Q_1^2 - Q_2^2$ and $2Q_1Q_2$ are transformed according to the representation $\Gamma_5$, as well \cite{Sofia13}. Moreover, these functions forming the basis of the representation $\Gamma_5$ are mutually transformed one into another under the action of the D$_\textrm{3d}$ space group symmetry elements. In order to obtain a result of the action of a D$_\textrm{3d}$ group element on the first of the above functions, one should utilize the first row of $\Gamma_5$ matrix, while in the case of the second function, the second row. Finally, the resulting $D(Q_1, Q_2)$ matrix for the vibronic coupling  $\Gamma_5$--$\Gamma_5$, or $\Gamma_5$--$\Gamma_6$ can be written as a sum of invariants,
\begin{equation}\label{eq12}
D(Q_1,Q_2)=\frac12\omega^2\left(Q_1^2+Q_2^2\right)\,\sigma_1+VQ_1\sigma_x+ W\frac12 W Q_1 Q_2
 \sigma_x+VQ_2\sigma_z+W\left(Q_1^2-Q_2^2\right)\sigma_z\,,
\end{equation}
where
$\omega$ denotes the frequency of a phonon in the harmonic approximation, $V$ and $W$ are matrix elements of the linear and quadratic vibronic coupling, $\Gamma_1$, and $\Gamma_6$ denote representations which describe degenerate electron states of the CuInP$_2$S$_6$ protostructure. The solution of the secular equation
\begin{equation}\label{eq13}
\left| D(Q_1,Q_2)-\varepsilon\right|=0,
\end{equation}
leads to the adiabatic potential.

\begin{figure}[!t]
\centerline{
\includegraphics[width=0.33\textwidth]{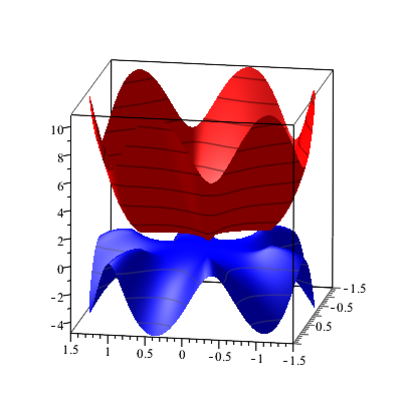}
\includegraphics[width=0.33\textwidth]{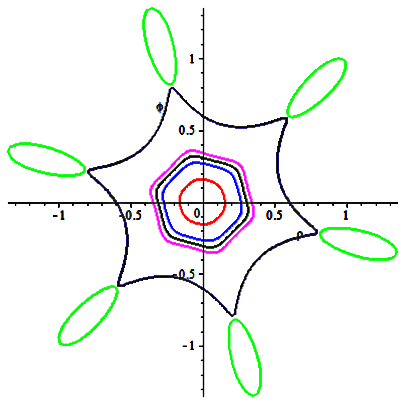}
\includegraphics[width=0.33\textwidth]{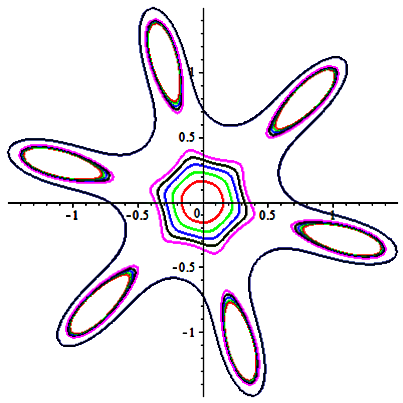}
}
\caption{(Color online) Left: A dependence of the adiabatic potential $\varepsilon_{1,2}$ of CuInP$_2$S$_6$ protostructure versus polar coordinates $\rho$ and $\varphi$. Middle and right: cross sections of the adiabatic potential for various energy values marked by different colors.}\label{fig7}
\end{figure}

In order to obtain a more convenient expression for further manipulations, one shall express equation (\ref{eq12}) in the Cartesian coordinates. It should be taken into account at the same time that the basis of representation $\Gamma_5$ are the $x^2-y^2$ and $2xy$ functions. Next, changing to the polar coordinate system $(\rho,\varphi)$ in which the $\rho$ axis coincides with the $\bf c$ leading axis of the protostructure crystal, we are in position to write the $D(\rho, \varphi)$ matrix elements as follows:
\begin{equation}\label{eq14}
\begin{split}
{{D}_{11}}&=\frac{1}{2}{{\omega }^{2}}{{\rho }^{4}}+V{{\rho }^{2}}\sin 2\varphi +W{{\rho }^{4}}\cos 4\varphi,  \\
{{D}_{22}}&=\frac{1}{2}{{\omega }^{2}}{{\rho }^{4}}-V{{\rho }^{2}}\sin 2\varphi -W{{\rho }^{4}}\cos 4\varphi,  \\
D_{12}&=D_{21}=V\rho^2\cos2\varphi +W\rho^4 \sin 4\varphi.
\end{split}
\end{equation}
The adiabatic potential $\varepsilon$ calculated from the respective secular equation takes the form
\begin{equation}\label{eq15}
{{\varepsilon }_{1,2}}\left( \rho ,\varphi  \right)=\frac{{{\omega }^{2}}{{\rho }^{4}}}{2}\pm {{\left[ V^2{{\rho }^{4}}+2VW{{\rho }^{6}}\sin 6\varphi +{{W}^{2}}{{\rho }^{8}} \right]}^{1/2}}.
\end{equation}
A dependence of the adiabatic potential versus $\rho$ and $\varphi$ coordinates, together with its cross-sections for various energy values, is presented in figure~\ref{fig7}.

As can be seen from figure~\ref{fig7}, the obtained adiabatic potential possesses 6 minima. Such a shape of the adiabatic potential allows a transition from the CuInP$_2$S$_6$ protostructure to the paraphase of the symmetry C$_\textrm{2h}^6$, since its symmetry with respect to the inversion operation is then preserved.

It should be noted here that the symmetry of the CuInP$_2$S$_6$ protostructure crystal is the same as that of paraphase of a related CuInP$_2$Se$_6$ crystal. The shape of adiabatic potential (\ref{eq15}) for the CuInP$_2$Se$_6$ paraphase allows, therefore, its phase transition to the ferrielectric phase of the symmetry C$_\textrm{3v}^2$, at the simultaneous loss of symmetry with respect to the inversion operation. Moreover, the shape of (\ref{eq15}) reflects the presence of the inversion element in the symmetry group D$_\textrm{3d}^2$ of the CuInP$_2$S$_6$ protostructure. The $\varepsilon(\rho, \varphi)$ function which exhibits 6 minima is transformed during the phase transition to the paraelectric phase of CuInP$_2$S$_6$ to a function having 2 minima, whereas at the phase transition paraphase--ferriphase of CuInP$_2$Se$_6$, this function attains 3 minima.

The two-minima adiabatic potential can be obtained in an analogous way for the paraphase of \linebreak CuInP$_2$S$_6$. It is responsible for the realization of the Jahn-Teller pseudoeffect, which is connected with the existence of two nearby-in-energy electron states. As has been demonstrated in section~\ref{par_AbInitio}, when passing from the model protostructure of the CuInP$_2$S$_6$ crystal to its paraphase, there occurs a splitting of the double degenerate electron states in points $\Gamma$, $K$, and $H$ of the elementary energy band that forms the vicinity of the forbidden band gap (figure~\ref{fig4}). Moreover, two doubly degenerate electron states which are described in the $\Gamma$ point by representations $\Gamma_5$ and $\Gamma_6$ are reduced to representations of the monoclinic paraphase as follows: $\Gamma_5 \rightarrow \Gamma_2 +\Gamma_4$, $\Gamma_6 \rightarrow \Gamma_1 +\Gamma_3$. As a result, there occurs a splitting of two double degenerate states into four energy states, described by one-dimensional representations with different parity. The adiabatic potential takes the form:
\begin{equation}\label{eq16}
\varepsilon \left( Q \right)=\frac12{{Q}^{2}}\pm \sqrt{{{\Delta }^{2}}+{{V}^{2}}{{Q}^{2}}}\,.
\end{equation}
Due to the presence of pairs of states $\Gamma_1$ and $\Gamma_2$ as well as $\Gamma_3$ and $\Gamma_4$ with the opposite parity, the Jahn-Teller pseudoeffect can take place in the system. States with the opposite parity are connected by a phonon with odd parity. This leads to the appearance of the dipole moments in the CuInP$_2$S$_6$ crystal. As a consequence of the vibronic interaction between the quasi-degenerate electronic state and a polar phonon, the levels nearby-in-energy become apart from one another, a connection between them is lost, and the adiabatic potential becomes a single minimum one. In the case of double minimum adiabatic potential related to the CuInP$_2$S$_6$ paraphase, two possible atomic sites Cu$^\textrm{u}$ and Cu$^\textrm{d}$ of copper in the unit cell are filled with equal probability. When the transition to the single minimum adiabatic potential takes place, the ordering of dipoles occurs with the appearance of spontaneous polarization.

\section{Conclusions}\label{par_Concl}

\vspace{-1mm}

\looseness=-1It has been demonstrated in this paper that it is enough to use a local symmetry of the crystal's actual Wyckoff position where the Jahn-Teller center is located in order to obtain a correct theory of the order-disorder type phase transitions in crystals. The group-theoretical approach together with the Pikus' method of invariants can be successfully utilized to construct the vibronic potential energy, and the adiabatic potential of a crystal. At the same time, it should be taken into account that the information on the symmetry and topology of the whole energy band structure of a crystal is encoded in the symmetry of the actual Wyckoff position. The developed approach allows one to find the mechanisms of the effect on the parameters of the phase transition in the CuInP$_2$S$_6$ crystal, which will be the subject of the study elsewhere.

\section*{Acknowledgements}%

\vspace{-1mm}

The authors would like to thank Prof.~Kharkhalis~L.Yu. for valuable
discussions and her interest to the work.

\appendix
\section{Tables of irreducible representations}\label{App}

\vspace{-5mm}


\begin{table}[!h]
\caption{Characters of irreducible representations of the D$_\textrm{3d}$ point group, as well as of the wave vector group in ${\bf k}=0$ for the D$_\textrm{3d}^2$ space group (notation of symmetry elements and representations in tables~\ref{tabA1}--\ref{tabA3} is adapted from reference \cite{Kovalev}. $h_{13}$ denotes inversion element).}
\label{tabA1}
\vspace{2ex}
\begin{center}
\renewcommand{\arraystretch}{0}
\begin{tabular}{|c|c|c|c|c|c|c|}
\hline
   &	\;\; $h_1$ \;\;&	 $h_3$, $h_5$ \;\; &	 $h_8$, $h_{10}$, $h_{12}$ \;\; &	 $h_{13}$ \;\;&	 $h_{15}$, $h_{17}$ \;\; &	 $h_{20}$, $h_{22}$, $h_{24}$ \strut\\ \hline\hline
 $\Gamma_1$ ($A_g$) &	1&	 1&	 1&	 1&	 1&	 1 \strut\\
 $\Gamma_2$ ($A_u$) &	1&	 1&	 1&	--1&	--1& --1 \strut\\
 $\Gamma_3$ ($B_g$) &	1&	 1&	--1&	 1&	 1& --1 \strut\\
 $\Gamma_4$ ($B_u$) &	1&	 1&	--1&	--1&	--1&  1 \strut\\
 $\Gamma_5$ ($E_g$) &	2&	--1&	 0&  2&	--1&  0 \strut\\
 $\Gamma_6$ ($E_u$) &	2&	--1&	 0&	--2&  1&  0 \strut\\ \hline
\end{tabular}
\renewcommand{\arraystretch}{1}
\end{center}
\end{table}
\vspace{-4mm}
\begin{table}[!h]
\caption{Characters of the irreducible representations of the C$_\textrm{2h}$ point group, as well as of the wave vector group in ${\bf k}=0$ for the C$_\textrm{2h}^5$ space group. ($h_{25}$ denotes inversion element).}
\label{tabA2}
\vspace{2ex}
\begin{center}
\renewcommand{\arraystretch}{0}
\begin{tabular}{|c|c|c|c|c|}
\hline
   &	\; $h_1$ \; &	 $h_4$ \; &	 $h_{25}$ \; &	 $h_{28}$ \strut\\ \hline\hline
 $\Gamma_1$ ($A_g$) &	1&	 1&	 1&	 1 \strut\\
 $\Gamma_2$ ($A_u$) &	1&	 1&	--1& --1 \strut\\
 $\Gamma_3$ ($B_g$) &	1&	--1&	 1&	--1 \strut\\
 $\Gamma_4$ ($B_u$) &	1&	--1&	--1&	 1 \strut\\ \hline
\end{tabular}
\renewcommand{\arraystretch}{1}
\end{center}
\end{table}

\begin{table}[!t]
\centering
\caption{Irreducible representations $\Gamma_5$ ($E_g$) and $\Gamma_6$ ($E_u$), written as real matrices (in cartesian coordinates).}
\label{tabA3}
\vspace{2ex}
\begin{center}
\renewcommand{\arraystretch}{0}
\begin{tabular}{|c|c|c|c|c|c|c|}
\hline
& \;\; $h_1$ \;\; &\;\; $h_3$ \;\; &\;\; $h_5$ \;\; &\;\; $h_8$ \;\; &\;\; $h_{10}$ \;\; &\;\; $h_{12}$ \strut\\ \hline\hline
$\Gamma_5$ &
   $\begin{pmatrix}
     1 & 0 \\[1em]
     0 & 1 \\
   \end{pmatrix}$ &
   $\begin{pmatrix}
   -\frac12 & -\frac{\sqrt3}{2} \\[1em]
     \frac{\sqrt3}{2} & -\frac12 \\
   \end{pmatrix}$ &
   $\begin{pmatrix}
   -\frac12 & \frac{\sqrt3}{2} \\[1em]
   -\frac{\sqrt3}{2} & -\frac12 \\
   \end{pmatrix}$ &
    $\begin{pmatrix}
     1 & 0 \\[1em]
     0 &-1 \\
   \end{pmatrix}$ &
   $\begin{pmatrix}
   -\frac12 & \frac{\sqrt3}{2} \\[1em]
    \frac{\sqrt3}{2} & \frac12 \\
   \end{pmatrix}$ &
    $\begin{pmatrix}
   -\frac12 & \frac{\sqrt3}{2} \\[1em]
    \frac{\sqrt3}{2} & \frac12 \\
   \end{pmatrix}$  \strut \\ 
  &\;\; $h_{13}$\;\; &\;\; $h_{15}$ \;\; &\;\; $h_{17}$ \;\; &\;\; $h_{20}$ \;\; &\;\; $h_{22}$ \;\; & $h_{24}$ \strut\\ \hline
$\Gamma_5$ &
   $\begin{pmatrix}
     1 & 0 \\[1em]
     0 & 1 \\
   \end{pmatrix}$ &
   $\begin{pmatrix}
   -\frac12 & -\frac{\sqrt3}{2} \\[1em]
     \frac{\sqrt3}{2} & -\frac12 \\
   \end{pmatrix}$ &
   $\begin{pmatrix}
   -\frac12 & \frac{\sqrt3}{2} \\[1em]
   -\frac{\sqrt3}{2} & -\frac12 \\
   \end{pmatrix}$ &
    $\begin{pmatrix}
     1 & 0 \\[1em]
     0 &-1 \\
   \end{pmatrix}$ &
   $\begin{pmatrix}
   -\frac12 & \frac{\sqrt3}{2} \\[1em]
    \frac{\sqrt3}{2} & \frac12 \\
   \end{pmatrix}$ &
    $\begin{pmatrix}
   -\frac12 & \frac{\sqrt3}{2} \\[1em]
    \frac{\sqrt3}{2} & \frac12 \\
   \end{pmatrix}$  \strut \\ \hline\hline
& \;\; $h_1$ \;\; &\;\; $h_3$ \;\; &\;\; $h_5$ \;\; &\;\; $h_8$ \;\; &\;\; $h_{10}$ \;\; &\;\; $h_{12}$ \strut\\ \hline\hline
$\Gamma_6$ &
   $\begin{pmatrix}
     1 & 0 \\[1em]
     0 & 1 \\
   \end{pmatrix}$ &
   $\begin{pmatrix}
   -\frac12 & -\frac{\sqrt3}{2} \\[1em]
     \frac{\sqrt3}{2} & -\frac12 \\
   \end{pmatrix}$ &
   $\begin{pmatrix}
   -\frac12 & \frac{\sqrt3}{2} \\[1em]
   -\frac{\sqrt3}{2} & -\frac12 \\
   \end{pmatrix}$ &
    $\begin{pmatrix}
     1 & 0 \\[1em]
     0 &-1 \\
   \end{pmatrix}$ &
   $\begin{pmatrix}
   -\frac12 & \frac{\sqrt3}{2} \\[1em]
    \frac{\sqrt3}{2} & \frac12 \\
   \end{pmatrix}$ &
    $\begin{pmatrix}
   -\frac12 & \frac{\sqrt3}{2} \\[1em]
    \frac{\sqrt3}{2} & \frac12 \\
   \end{pmatrix}$  \strut \\ 
  & \;\; $h_{13}$ \;\; &\;\; $h_{15}$ \;\; &\;\; $h_{17}$ \;\; &\;\; $h_{20}$ \;\; &\;\; $h_{22}$ \;\; & $h_{24}$ \strut\\ \hline
$\Gamma_6$ &
   $\begin{pmatrix}
     -1 & 0 \\[1em]
     0 & -1 \\
   \end{pmatrix}$ &
   $\begin{pmatrix}
   \frac12 & \frac{\sqrt3}{2} \\[1em]
   -\frac{\sqrt3}{2} & \frac12 \\
   \end{pmatrix}$ &
   $\begin{pmatrix}
   \frac12 & -\frac{\sqrt3}{2} \\[1em]
   \frac{\sqrt3}{2} & \frac12 \\
   \end{pmatrix}$ &
    $\begin{pmatrix}
     -1 & 0 \\[1em]
     0 & 1 \\
   \end{pmatrix}$ &
   $\begin{pmatrix}
   \frac12 & -\frac{\sqrt3}{2} \\[1em]
   -\frac{\sqrt3}{2} & -\frac12 \\
   \end{pmatrix}$ &
    $\begin{pmatrix}
   \frac12 &-\frac{\sqrt3}{2} \\[1em]
    -\frac{\sqrt3}{2} & -\frac12 \\
   \end{pmatrix}$  \strut \\ \hline
\end{tabular}
\renewcommand{\arraystretch}{1}
\end{center}
\end{table}





\ukrainianpart

\title{Вібронна взаємодія в кристалах з ян-теллеровськими центрами в концепції мінімальних комплексів зон}
\author{Д.М. Берча\refaddr{a1}, С.А. Берча\refaddr{a1}, К.Є. Глухов\refaddr{a1}, М. Шнайдер\refaddr{a2}}
\addresses{
\addr{a1} Інститут фізики та хімії твердого тіла, Ужгородський національний університет,
вул. Волошина 54, 88000 Ужгород, Україна
\addr{a2}Факультет математичних та природничих наук, Університет Жешува, вул. Пігонія 1, 35-959
Жешув, Польща
}
\makeukrtitle
\begin{abstract}
\tolerance=3000%
Аналізується фазовий перехід типу порядок-безпорядок    в моноклінному
 кристалі CuInP$_2$S$_6$,  спричинений вібронними взаємодіями (ефект
 Яна-Телера). З цією метою
 створено модель тригональної протостроктури для  CuInP$_2$S$_6$  шляхом
 невеликої зміни  параметрів гратки CuInP$_2$S$_6$  в параелектричній фазі.
 Одночасно з теоретикогруповим аналізом, здійснюється першопринципний
 розрахунок на основі методу функціоналу густини зонної структури CuInP$_2$S$_6$
 в протоструктурі, пара- і ферофазах. Використовуючи концепцію елементарних
 енергетичних зон, встановлено зв'язок частини зонної структури
 в околі забороненої енергетичної зони, що створюється  станами  $d$-електронів
 міді, з певним  положенням Вікоффа, де локалізовані ян-телерівські
 центри. Процедура побудови матриці вібронної потенціальної взаємодії
 узагальнюється на випадок кристалу, використовуючи
 концепцію елементарних енергетичних зон і теоретикогрупового методу
 інваріантів. Процедура ілюструється на прикладі створення
 адіабатичних потенціалів вібронного зв'язку $\Gamma_5$--$\Gamma_5$ для
 протоструктури і парафази кристалу  CuInP$_2$S$_6$. На основі аналізу отриманої
 структури
 адіабатичних потенціалів  зроблено висновки щодо їх перетворення при фазовому
 переході і обговорено можливість виникнення в
 кристалі спонтанної поляризації.
 \keywords ефект Яна-Телера, адіабатичні потенціали, положення Вікоффа, теорія груп
\end{abstract}

\end{document}